\documentclass[prd,twocolumn]{revtex4}
\usepackage{graphicx, epsfig}
\usepackage{color}
\usepackage{mathrsfs}
\usepackage{bm}

\newcommand{\be}{\begin{equation}}
\newcommand{\ee}{\end{equation}}
\newcommand{\bea}{\begin{eqnarray}}
\newcommand{\eea}{\end{eqnarray}}
\newcommand{\ba}{\begin{eqnarray}}
\newcommand{\ea}{\end{eqnarray}}

\newcommand{\gapp}{\mathrel{\raise.3ex\hbox{$>$}\mkern-14mu
              \lower0.6ex\hbox{$\sim$}}}
\newcommand{\lapp}{\mathrel{\raise.3ex\hbox{$<$}\mkern-14mu
              \lower0.6ex\hbox{$\sim$}}}

\begin{document}
\title{Aharonov-Bohm Radiation}

\author{Katherine Jones-Smith$^1$, Harsh Mathur$^1$, Tanmay Vachaspati$^{1,2}$} 
\affiliation{
$^1$CERCA, Department of Physics, 
Case Western Reserve University, Cleveland, OH~~44106-7079.\\
$^2$Institute for Advanced Study, Princeton, NJ 08540. 
}

\begin{abstract}
\noindent
A solenoid oscillating in vacuum will pair produce charged particles 
due to the Aharonov-Bohm (AB) interaction. We calculate the radiation 
pattern and power emitted for charged scalar particles. We extend the 
solenoid analysis to cosmic strings, and find enhanced radiation from 
cusps and kinks on loops. We argue by analogy with the electromagnetic 
AB interaction that cosmic strings should emit photons due to the 
gravitational AB interaction of fields in the conical spacetime of 
a cosmic string. We calculate the emission from a kink and find
that it is of similar order as emission from a cusp, but kinks are 
vastly more numerous than cusps and may provide a more interesting 
observational signature. 
\end{abstract}

\maketitle

Fifty years ago Aharonov-Bohm (AB) \cite{Aharonov:1959fk} 
showed that charged particles can scatter non-trivially in 
the pure gauge potential outside a thin solenoid. The effect 
has been experimentally investigated in various setups 
\cite{Yauetal2002,Tonomuraetal1986} and quite recently 
(theoretically) in graphene \cite{Jackiw:2009bw}. In cosmology, 
AB scattering is relevant to the dynamics of cosmic strings, 
as they move in the ambient cosmological medium 
\cite{Rohm1985,Alford:1988sj} and the AB interaction could be 
a link between dark strings and the visible sector 
\cite{Vachaspati:2009jx}.

AB scattering is purely quantum, having no classical analog, 
and also purely topological, as it arises from the non-trivial change 
in the phase of the wavefunction of a charged particle
as it is taken in a closed path around the solenoid. In this paper
we study if the AB interaction can also 
lead to particle creation -- if a thin long solenoid oscillates, 
does it produce electron-positron pairs? Such radiation would 
be purely quantum and purely topological.

The question of AB radiation has been partially addressed by 
Alford and Wilczek \cite{Alford:1988sj} where the problem is 
set up (but not solved). In this paper, we shall take
up this task using two approaches. The first is to use the AB 
phase, defined as $\epsilon = e \Phi$ where $e$ is the
charge and $\Phi$ the magnetic flux through the solenoid,
as a small control parameter. Here the calculation can be 
done using the Feynman 
diagram language with the gauge potential of the moving solenoid
treated as a classical background. The method can also be applied
to find the AB radiation rate from oscillating cosmic string 
loops. The disadvantage of this perturbative approach is that 
we lose track of the expected periodicity of all quantities in 
the AB phase. For example, the classic ``$\sin (\pi \alpha)$''
behavior of AB scattering, where $\alpha = \epsilon /2\pi$,
does not appear in this treatment.
Our second approach is to allow the AB phase to take any value 
but to only consider slowly oscillating solenoids {\it i.e.} we
expand in the velocity of the solenoid. The advantage now is 
that we explicitly see the periodic dependence of the radiation 
rate on the AB phase. However, we cannot treat relativistic motion 
such as that of cosmic string loops in this approximation.
We can, of course, take both the AB phase and the oscillation 
speed to be small, in which case both methods should agree, 
and we show that they do.

Another novel aspect of AB radiation is that it may also apply
in the gravitational context. In a conical metric, such as that
of a cosmic string, the wavefunction of any particle that goes
around the string acquires a phase equal to the conical deficit,
which we denote $\delta = 8 \pi G\mu$ where $G$ is Newton's
gravitational constant and $\mu$ is the string tension. Then,
if a string oscillates, it emits all particles including,
for example, photons. This may make cosmic strings, even those
whose constituent fields have no interactions with photons, 
visible in the electromagnetic domain. We will discuss 
gravitational AB radiation in Sec.~\ref{gravAB} where we
rely on the analysis developed in the earlier sections.
Although we arrived at gravitational AB radiation via
the gauge AB process, the effect is just that of particle
production in the time-dependent gravitational background 
of a cosmic string considered by Garriga, Harari and
Verdaguer in \cite{Garriga:1989bx} (also see
\cite{Frieman:1985fr}).

The rest of the paper is organized so that we first discuss the
gauge potential produced by a moving solenoid in Sec.~\ref{gaugepot},
as given in \cite{Alford:1988sj}. Then we find the AB radiation
from an infinite, straight, oscillating solenoid in Sec.~\ref{solenoidAB}, 
using first the small AB phase approximation and then in the slow 
velocity approximation. In Sec.~\ref{loopsAB} we
calculate AB radiation from cosmic string loops, providing two
explicit examples, one of a loop with kinks and the other of a loop 
with cusps but no kinks. In Sec.~\ref{gravAB} we discuss gravitational 
AB radiation. We conclude in Sec.~\ref{conclusions}. The formalism
to find AB radiation in slowly changing backgrounds is set up
in Appendix~\ref{appendixframe}.

In this paper we have restricted ourselves to AB radiation of 
spin zero (scalar) particles and are mostly concerned with the
case when the emitted particles are massless. The extension to
fermions is conceptually similar but the spin orientation 
provides another degree of freedom and makes the calculations
more involved at a technical level. We will address the fermion 
calculation in a separate publication \cite{Yi-ZenChu}.

\section{Gauge Potential of Moving Solenoid}
\label{gaugepot}

The first step in calculating AB radiation is to find the gauge 
potential for a moving solenoid. This step was taken in Alford and 
Wilczek \cite{Alford:1988sj}. 

A moving solenoid with magnetic flux $\Phi$ will be a current source 
in Maxwell's equation, and the current can be written as
\begin{equation}
J^{(\Phi )}_\nu = 
\frac{\Phi}{2} \epsilon_{\mu\nu\alpha\beta} \partial^\mu S^{\alpha\beta} 
\end{equation}
where
\begin{eqnarray}
S^{\alpha\beta}(x) &=& \int d\tau d\sigma \sqrt{-\gamma}
     \epsilon^{ab} ~ \partial_a  X^\alpha \partial_b X^\beta 
                 \delta^{(4)} (x-X(\sigma,\tau))
\nonumber \\
&& \hskip -1 cm = \int d\tau d\sigma 
     ( {\dot X}^\alpha {X^\beta} ' - {\dot X}^\beta {X^\alpha} ' )
                 \delta^{(4)} (x-X(\sigma,\tau))
\end{eqnarray}
where $\sigma , \tau$ are world-sheet coordinates, $X^\mu (\sigma)$ 
denotes the position of the string, overdots and primes denote
derivatives with respect to $\tau$ and $\sigma$ respectively,
and $\gamma_{ab}$ is the worldsheet metric, with $\gamma$
denoting its determinant. 
In what follows, we shall use $\tau =t$, while $\sigma =z$ for
a solenoid along the $z-$axis. In applications to cosmic
strings, we will also use the gauge conditions
\begin{equation}
{\dot X} \cdot X' = 0 \ , \ \ 
{\dot X}^2 + {X'}^2 =0
\label{gaugeconditions}
\end{equation}

The field strength of a static, thin solenoid satisfies
\begin{equation}
\partial^\mu F_{\mu\nu} = J_\nu
\end{equation}
and the solution is
\begin{equation}
F_{\mu\nu} = \frac{\Phi}{2} 
           \epsilon_{\mu\nu\alpha\beta} S^{\alpha\beta}
\label{solenoidF}
\end{equation}
The corresponding gauge potential can be written in
Lorenz gauge
\begin{equation}
A_\nu = \frac{\Phi}{2} \epsilon_{\mu\nu\alpha\beta} \partial^\mu 
       \frac{1}{\partial^2} S^{\alpha\beta}
\end{equation}
where $\partial^2$ is the D'Alembertian operator. We will only
need the gauge potential in momentum space,
\begin{equation}
{\tilde A}_\nu = - i \frac{\Phi}{2} \epsilon_{\mu\nu\alpha\beta} 
       \frac{k^\mu}{k^2} {\tilde S}^{\alpha\beta}
\end{equation}
where overtilde's denote the Fourier transformed variable
\begin{equation}
{\tilde S}^{\alpha\beta} (k) =
\int d^4x ~ e^{+ik\cdot x} S^{\alpha\beta}(x) \ .
\end{equation}

Note from Eq.~(\ref{solenoidF}) that the field strength vanishes 
everywhere except at the location of the solenoid and so a 
moving solenoid does not radiate classical electromagnetic waves.

\section{AB Radiation}
\label{ABradiation}

\begin{figure}
\scalebox{1.0}{\includegraphics[angle=0]{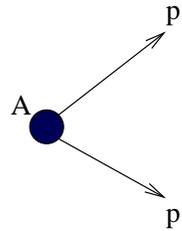}}
\caption{Feynman diagram for pair creation by oscillating solenoid.
The solenoid creates an oscillating gauge potential, $A_\mu$, while 
the field strength is always zero outside the solenoid. The time
dependent pure gauge $A_\mu$ leads to particle production.
}
\label{feynmandiag}
\end{figure}

Consider the interaction term $J_\mu A^\mu$ where, for
example, for a complex scalar field, $\chi$, we have
\begin{equation}
J_\mu = ie (\chi^* \partial_\mu \chi - 
            \chi \partial_\mu\chi^* )
\end{equation}
where $e$ is the charge of $\chi$. 
The lowest order amplitude for $\chi$ production is found
from the diagram in Fig.~\ref{feynmandiag} and can be written
as
\begin{equation}
i{\cal M} = -i \frac{\Phi}{2} 
 \epsilon_{\mu\nu\alpha\beta} \frac{k^\mu}{k^2}
 {\tilde S}^{\alpha\beta} (k) J^\nu (p,p') |_{k=p+p'}
\end{equation}
with $J^\nu (p,p') = ie(p-p')^\nu$.
Then
\begin{eqnarray}
i{\cal M} &=& \frac{\epsilon}{2} ~ \epsilon_{\mu\nu\alpha\beta} 
             \frac{(p+p')^\mu (p-p')^\nu}{(p+p')^2} \nonumber \\
     &\times& \int d\tau d\sigma 
({\dot X}^\alpha {X^\beta}' - {\dot X}^\beta {X^\alpha}')
                 e^{-i (p+p') \cdot X} 
\label{amplitude}
\end{eqnarray}
where $\epsilon = e\Phi$ is the AB phase and is assumed to 
be small.

The number of particles produced is given by (see, for
example, Eq.~(4.74) of \cite{PeskinSchroeder})
\begin{equation}
dN = \frac{d^3p}{(2\pi)^3} \frac{1}{2\omega} 
     \frac{d^3p'}{(2\pi)^3} \frac{1}{2\omega'} |{\cal M}|^2
\label{eq:dN}
\end{equation}
where $\omega = \sqrt{{\bm p}^2 + m^2}$ (similarly $\omega'$)
and $m$ is the mass of the scalar particle.

\section{AB Radiation from Oscillating Solenoid}
\label{solenoidAB}

In this section we will consider AB radiation from an infinite straight 
solenoid that is oscillating back and forth. We will do the calculation 
in the small AB phase approximation. The oscillating solenoid is also
treated in the non-relativistic approximation in subsection~\ref{smallfreq}, 
and the results of the two methods agree in the regime where both are valid.

We will take the position of the solenoid to be described by
\begin{equation}
X^\mu = (t,A \sin (\Omega t), 0, \sigma )
\label{solenoidX}
\end{equation}
where $A$ is the amplitude and $\Omega$ the frequency of
oscillation. Note that $X^\mu$ is not the solution to any
equation of motion but just a convenient oscillatory choice
that provides us with a background in which to study AB radiation.

\subsection{Small AB Phase Approximation}
\label{smallphase}

Inserting Eq.~(\ref{solenoidX}) in Eq.~(\ref{amplitude}) gives
\begin{equation}
i{\cal M} = -\frac{2\epsilon}{(p+p')^2} 
 \left [ (p_x p_y'-p_y p_x')I_0 + (\omega 'p_y - \omega p_y') I_1 \right ]
\label{iM1}
\end{equation}
where $\omega = p^0$, $\omega'={p^0}'$,
\begin{equation}
I^0 = \int d\tau d\sigma ~ e^{-i (p+p')\cdot X}
\end{equation}
\begin{equation}
I^1 = \int d\tau d\sigma ~ \Omega A \cos(\Omega \tau) 
e^{-i (p+p')\cdot X}.
\end{equation}
The integration along the straight string is trivial, and leads 
to a $2\pi \delta (p_z+p_z')$ factor in $I^0$ and $I^1$.

To simplify further, we use the relation
\begin{equation}
e^{i\alpha \sin(\Omega\tau)} = 
      \sum_{n=-\infty}^{+\infty} J_n (\alpha) e^{in\Omega\tau}
\end{equation}
which can be checked using the integral representation for the
Bessel functions
\begin{equation}
J_n (\alpha) = \frac{1}{2\pi} \int_{-\pi}^{+\pi} 
                  e^{-i (n\tau -\alpha \sin\tau)} d\tau
\end{equation}
Then
\begin{equation}
I^0 = \frac{(2\pi)^2}{\Omega} \sum_n J_n (P^1 A) \delta (n- (P^0/\Omega))
          \delta (p_z+p_z')
\end{equation}
and
\begin{eqnarray}
I^1 &=& 
2\pi^2 A \sum_n [ J_{n-1}(P^1A) + J_{n+1} (P^1 A)] \nonumber \\
    &&     \times \delta (n- (P^0/\Omega)) \delta (p_z+p_z')
\nonumber \\
&& \hskip -1 cm
= \frac{(2\pi)^2}{P^1} \sum_n n J_{n}(P^1A) \delta (n- (P^0/\Omega))
          \delta (p_z+p_z')
\end{eqnarray}
where $P = p+p'$ and $J_n (x)$ is the Bessel function of $n^{\rm th}$
order. 
                       
We can now insert the expressions for the integrals
$I^0$ and $I^1$ in Eq.~(\ref{iM1}) to get
\begin{eqnarray}
i{\cal M} &=& -8\pi^2 \epsilon \frac{(p_y-p_y')}{(p_x+p_x')} \nonumber \\
&&\hskip -1 cm \times \sum_{n=1}^\infty
       J_n((p_x+p_x')A) \delta (\omega+\omega'-n\Omega)
          \delta (p_z+p_z')
\label{amp}
\end{eqnarray}

\begin{figure}
\scalebox{0.25}{\includegraphics[angle=-90]{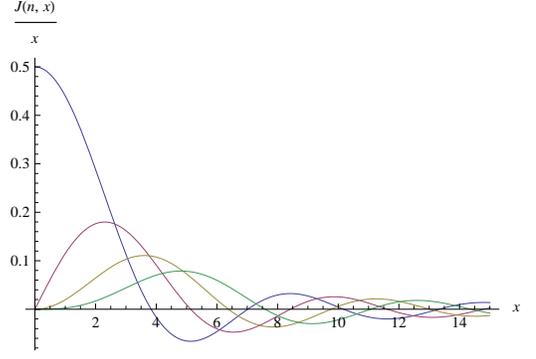}}
\caption{Plot of $J_n(x)/x$ for $n=1,2,3,4$, where the
amplitude of the peak decreases with increasing $n$.
}
\label{besseloverx}
\end{figure}

At first glance the amplitude depends on six variables, namely the six 
different momentum components. However, Eq.~(\ref{amp}) shows that it 
only depends on the combinations $p_x+p_x'$, $p_y-p_y'$ and $p_z+p_z'$ 
as long as the overall energy is a harmonic of the oscillation frequency. 
Note that the amplitude is not singular
at $p_x+p_x'=0$ because the Bessel functions are proportional 
to $(p_x+p_x')^n$ at small argument and the sum in Eq.~(\ref{amp}) 
starts at $n=1$ (see Fig.~\ref{besseloverx} for plots of 
$J_n(x)/x$). The amplitude is largest when $p_y-p_y'$ is 
maximum and $p_x+p_x'=0=p_z+p_z'$.
Together with the total energy constraint 
\begin{equation}
\omega +\omega ' =
\sqrt{{\bm p}^2+m^2} + \sqrt{{{\bm p}'}^2+m^2} = n\Omega ,
\end{equation}
we find that the maximum
amplitude is along the $y$ direction, with $p_x=0=p_x'$.

The number of particles emitted follows from Eq.~(\ref{eq:dN}).
On squaring ${\cal M}$ we get a double sum but the energy conservation
delta function reduces the expression to a single sum and introduces
a $\delta (0)$ on the right-hand side. Dividing out by $2\pi\delta (0)$
gives the {\it rate} of particle production, denoted by ${\dot N}$,
\begin{eqnarray}
d{\dot N} &=& \frac{d^3p}{(2\pi)^3} \frac{1}{2\omega} 
              \frac{d^3p'}{(2\pi)^3} \frac{1}{2\omega'} 
\frac{( 8\pi^2 \epsilon )^2}{2\pi} 
\left ( \frac{p_y-p_y'}{p_x+p_x'} \right )^2
     \nonumber \\
&&\hskip -1.8 cm \times \sum_{n=1}^\infty
       | J_n((p_x+p_x')A) |^2 \delta (\omega+\omega'-n\Omega) 
          \{ \delta (p_z+p_z') \}^2
\label{eq:dNdotsolenoid}
\end{eqnarray}
The integration over $p_z'$ leads to another factor of $\delta (0)$.
Dividing out by another factor of $2\pi \delta (0)$ gives
the rate of particle emission 
{\it per unit length} of the solenoid, denoted by ${\dot N}'$,
\begin{eqnarray}
d{\dot N}' &=& \frac{\epsilon^2}{16 \pi^4} dp_z d^2p_\perp d^2p_\perp'
 \frac{1}{\omega \omega '} \left ( \frac{p_y-p_y'}{p_x+p_x'} \right )^2
    \nonumber \\
  &\times &
 \sum_{n=1}^\infty | J_n((p_x+p_x')A) |^2 \delta (\omega+\omega'-n\Omega) 
\label{ddotNprime}
\end{eqnarray}
where the $\perp$ subscript denotes the $x$, $y$ components. Also,
now $\omega ' = \sqrt{{{\bm p}_\perp'}^2 + p_z^2 + m^2}$. 

The expression for the rate of particle emission depends on 
five of the momenta. The dependence on $p_z$ enters via 
$\omega$ and $\omega '$, and it is clear that the largest
emission is for $p_z =0$ {\it i.e.} in the plane perpendicular
to the solenoid. So we restrict our attention to $p_z =0$
and also to the massless case, $m=0$. These assumptions
imply that $\omega = |{\bm p}_\perp|$, $\omega' = |{\bm p}_\perp'|$,
and we obtain in polar coordinates
\begin{eqnarray}
\frac{d{\dot N}'}{dp_z} (p_z=0) &=& 
       \frac{\epsilon^2}{16\pi^4} dp_\perp dp_\perp'
          d\theta d\theta' \left ( \frac{p_y-p_y'}{p_x+p_x'} \right )^2
    \nonumber \\
  && \hskip -2 cm \times 
 \sum_{n=1}^\infty | J_n((p_x+p_x')A) |^2 \delta (p_\perp+p_\perp'-n\Omega) 
\label{dNperp}
\end{eqnarray}
where $p_x = p_\perp \cos \theta$ etc..

\begin{figure}
\scalebox{0.6}{\includegraphics{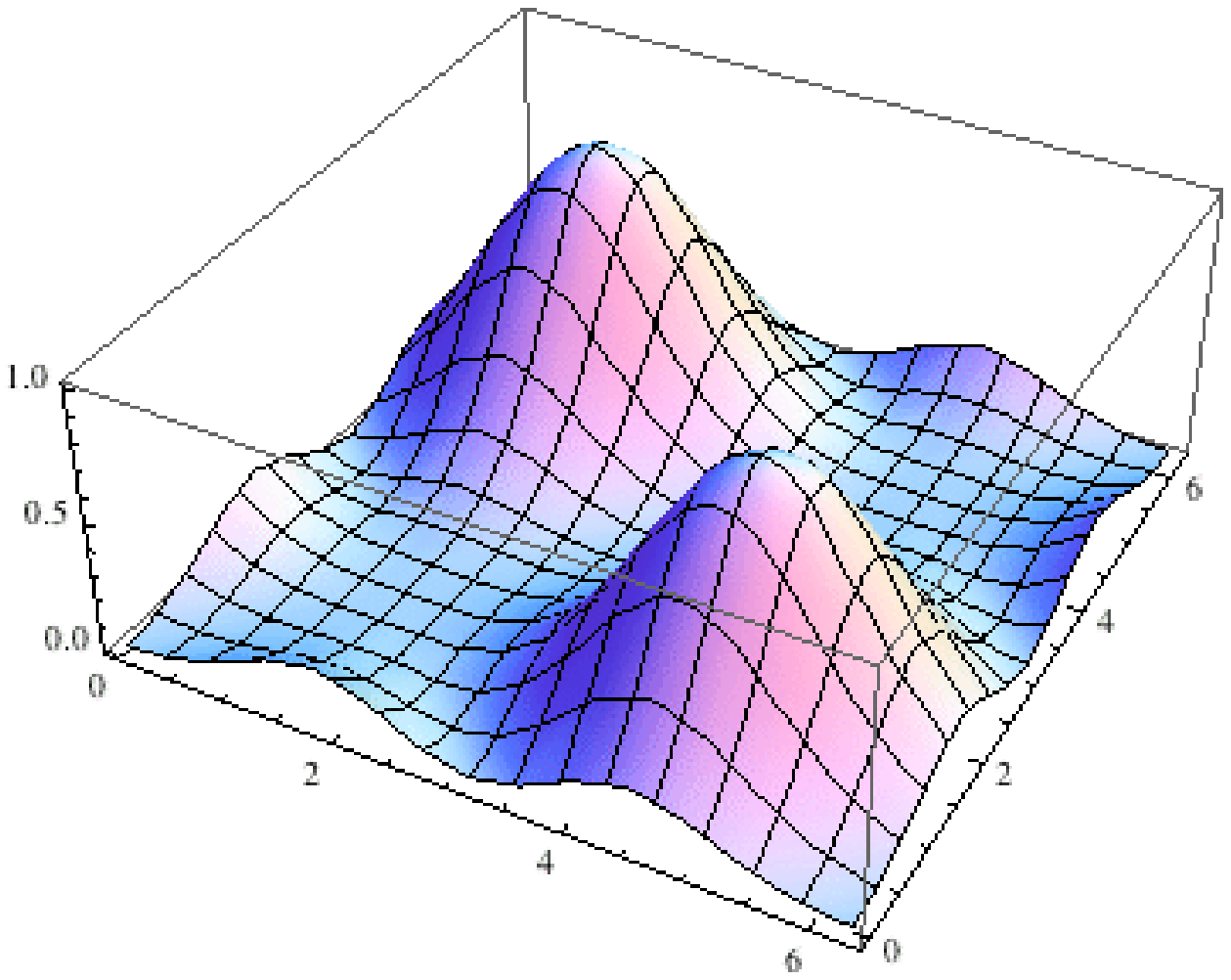}}
\scalebox{0.6}{\includegraphics{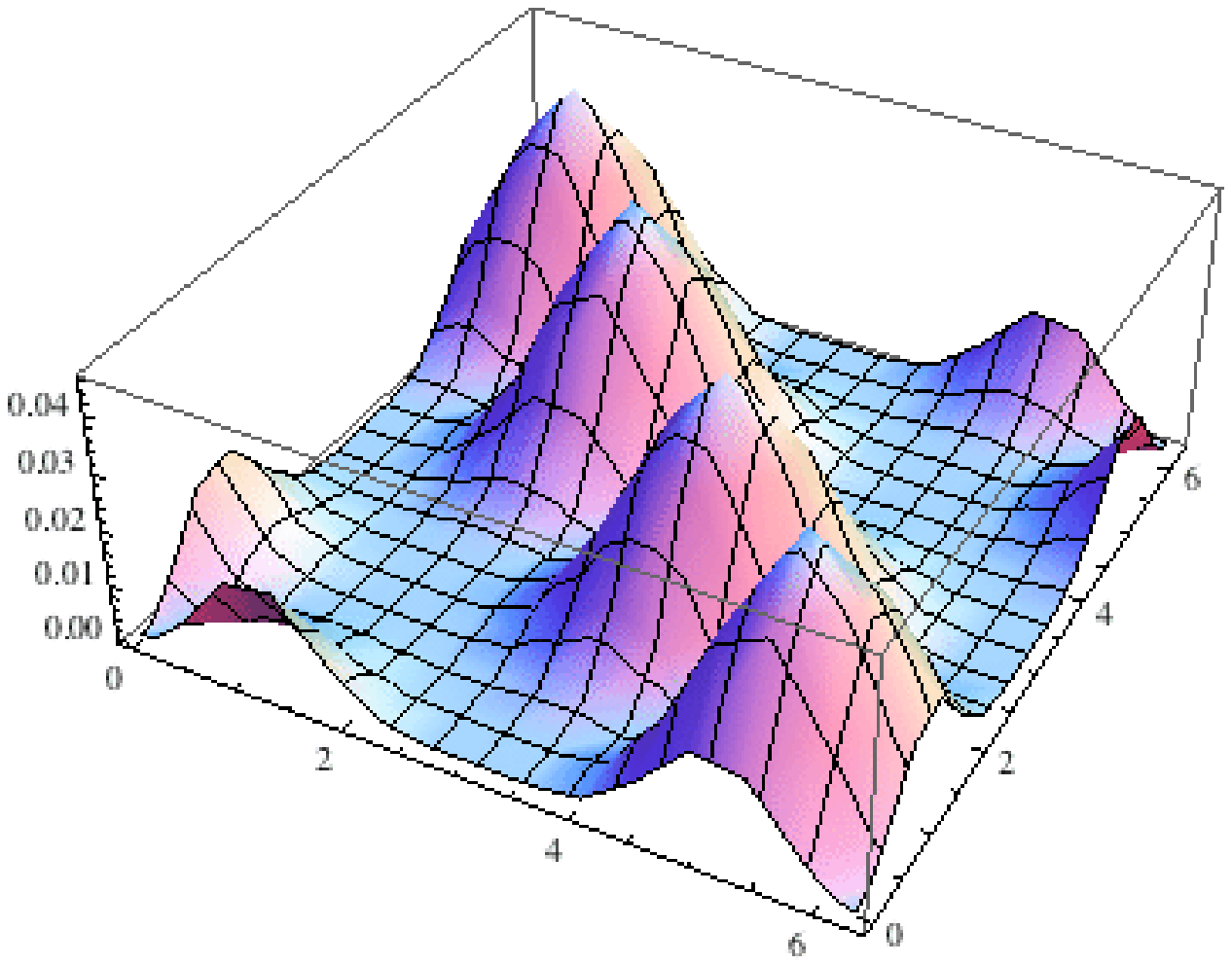}}
\caption{Here we show the radiated power as a function of $\theta$ 
and $\theta '$. The top figure is for $n=1$ with 
$|{\bm p}|=n\Omega/2 = |{\bm p}'|$ and the lower figure is for 
$n=2$. There is maximum radiation along the line 
$\theta +\theta ' =\pi$, implying back to back emission.
For $n=1$, the peaks occur at $\theta =\pi /2 , ~ 3\pi /2$. For 
$n=2$, they occur at $\theta =\pi /4 , ~ 3\pi /4, ~ 5\pi/4, ~ 7\pi/4$.
}
\label{angdistn}
\end{figure}

The integration over $p_\perp'$ can be done giving the emission
rate per unit length in the orthogonal plane as a function of the
three variables, $p_\perp$, $\theta$ and $\theta '$. The angular 
distribution is of greatest interest and is shown for $n=1$
and $n=2$ in Fig.~\ref{angdistn}, taking $p_\perp = n\Omega/2$.
The figure shows back to back emission, with the $n=1$ emission
being dipolar and maximum along the $y$ direction.

The total radiation rate can be found by integrating 
Eq.~(\ref{ddotNprime}) over momenta and also doing the sum
over harmonics. The sum is convergent because the Bessel functions 
at fixed argument fall off exponentially with $n$ for large $n$.
We will work only in the $m=0$ case and restrict attention to
particle emission in the orthogonal plane, as in Eq.~(\ref{dNperp}).
This is equivalent to working out the problem in two spatial
dimensions, and we shall denote the particle emission rate by
$N^{(2d)}$. The delta function can be integrated out by going to
polar coordinates. We get
\begin{equation}
{\dot N}^{(2d)} = \frac{\epsilon^2 \Omega }{16\pi^4 } 
                 \sum_{n=1}^\infty n C_n (a_n)
\end{equation}
where $a_n \equiv n\Omega A$ and
\begin{eqnarray}
C_n (a) &=& 
   \int_0^1 dq \int_{-\pi}^\pi d\theta \int_{-\pi}^{\pi} d\theta' 
       \left ( \frac{q \sin\theta - (1-q) \sin\theta'}
                   {q\cos\theta + (1-q) \cos\theta '} \right )^2
    \nonumber \\
  && \times | J_n (a(q\cos\theta + (1-q) \cos\theta ')) |^2 
\end{eqnarray}
These integrations have to be done numerically.

\begin{figure} 
\scalebox{0.33}{\includegraphics[angle=-90]{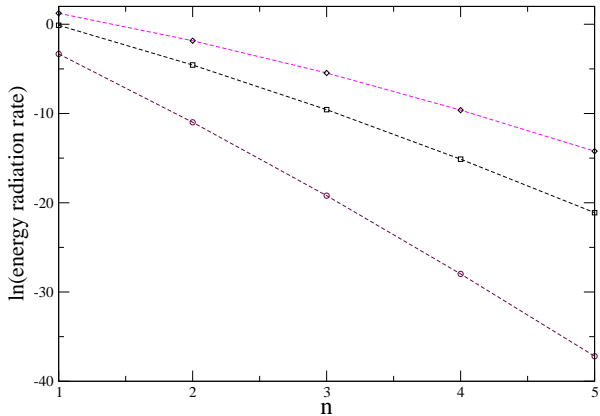}}    
\caption{$\ln({\dot {\cal E}}_n)$ vs. $n$ for $\Omega A =0.1$
(lowest curve), 0.5 (middle curve) and 1.0 (top curve).
}
\label{lnEvsn}
\end{figure}

The energy radiation rate in the $n^{th}$ harmonic is 
\begin{equation}
{\dot E}_n = n\Omega {\dot N}_n
\end{equation}
and Fig.~\ref{lnEvsn} shows a plot of $\ln({\dot {\cal E}}_n)$ 
versus $n$, where 
${\dot {\cal E}}_n = {\dot E}_n 16\pi^4/(\epsilon^2 \Omega^2)$ for
$\Omega A =0.1, 0.5, 1$. Note that the energy falls off rapidly
with increasing $n$ and so essentially all of the particles are 
emitted in the $n=1$ harmonic {\it i.e.} with energy $\Omega$.

\begin{figure}
\scalebox{0.33}{\includegraphics[angle=-90]{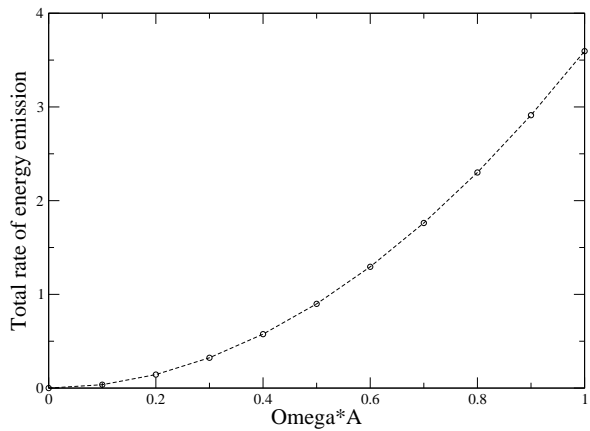}}
\caption{${\dot {\cal E}}$ vs. $\Omega A$ for the oscillating
solenoid.}
\label{totalEvsOmegaA}
\end{figure}

The total energy emission rate ${\dot {\cal E}}$ as a 
function of the parameter $\Omega A$ is shown in 
Fig.~\ref{totalEvsOmegaA}.

We have not calculated the emission rate for massive particles,
$m \ne 0$, in detail. From Eq.~(\ref{eq:dNdotsolenoid})
we see that $m$ enters
through $\omega$ and $\omega '$ in the phase space factor
and then again in the energy conservation delta function
but nowhere else. The delta function can be non-zero only
if $n \Omega > 2 m$ {\it i.e.} $n > 2m/\Omega$. If the
oscillation frequency is small compared to the mass of
the particle -- as is relevant for oscillating solenoids
in the laboratory when the charged particle is the
electron -- the Bessel functions decay exponentially 
with large $n$ and we expect the emission to be
exponentially suppressed. The exponential suppression
is also seen in the plot of Fig.~\ref{lnEvsn}.

It is noteworthy that the total energy of the pairs produced
add up to $n \Omega$ by virtue of the delta function constraint
in eq (24). Commonly in quantum mechanics, perturbations at
frequency $\Omega$ produce excitations with energy $\hbar \Omega$
(as for example in optical transitions in atomic physics). Here
excitations with energies equal to all multiples of the fundamental
frequency are produced (although in practice the higher harmonic
production may be small except in an ultra-relativistic limit).
This situation is also encountered in laboratory scale phenomena
such as Mossbauer spectroscopy and the excitation of conduction
electrons by the motion of impurities in metals. Mathematically
the appearance of excitations at higher harmonics is traceable to
the Fourier expansion eq (17).

In the next sub-section, using the formalism developed in
Appendix~\ref{appendixframe}, we re-analyze AB radiation from 
an oscillating solenoid for arbitrary values of the AB phase,
$\epsilon$, but in the limit of small $\Omega A$. We recover
the periodicity of the result as a function of AB phase.
We also show that the results of the two different methods 
agree in the limit that both $\epsilon$ and $\Omega A$ are 
small.

\subsection{Small Oscillation Frequency Approximation}
\label{smallfreq}

Now let us consider the circumstance that the flux is not a small 
perturbation. In this case we can still obtain analytic results 
for particle production in the limit that the solenoid motion 
is non-relativistic. These results are non-perturbative in the
flux and show the flux periodicity expected of the Aharonov-Bohm 
effect.  Moreover, the small flux limit of the results obtained in 
this section match the low velocity limit of the results in the 
previous section. 

The essential physics here is that for a stationary solenoid
the eigenmodes of the scalar field depend (periodically) on the 
flux $\Phi$ of the solenoid. If the solenoid is moved to a different 
location the eigenmodes and hence the quantum vacuum is altered. 
If the solenoid is moved slowly from the initial to the final 
configuration, in the adiabatic approximation it will simply pass from 
the initial vacuum to the final vacuum, but post-adiabatic corrections  
lead to particle production. The technical implementation of this
idea is relegated to Appendix~\ref{appendixframe} where we analyze 
the leading post-adiabatic dynamics of an assembly of oscillators 
with slowly varying eigenmodes. 

We will work in cylindrical co-ordinates and in the ``singular
gauge'' where the gauge potential of a static solenoid is
\begin{equation}
A^{\mu} = ( 0, 0, \Phi \Theta (x) \delta(y), 0 )
\label{eq:singular}
\end{equation}
Here $\Theta$ is the Heaviside function. The gauge potential
vanishes everywhere on the $xy-$plane and is singular along
the positive $x-$axis. It is easy to check that the line integral
of the gauge potential along a path enclosing the origin yields
the flux $\Phi$.

The eigenmodes, $\psi_{kql}$, for a scalar field in the background 
of a solenoid along the $z-$axis have been found in 
Ref.~\cite{Aharonov:1959fk}. We will use those eigenmodes in
singular gauge.
In order to box normalize the eigenmodes, we adopt periodic 
boundary conditions along the $z$-axis and take the length of the 
solenoid to be $L$. Along the radial direction we find it 
convenient to impose Dirichlet boundary conditions at $r=a$ 
({\it i.e.} $\psi_{kql} (a)=0$) if $l$ is even, and Neumann 
boundary conditions ({\it i.e.} $\partial_r \psi_{kql} (a)=0$)
if $l$ is odd. These boundary conditions are convenient because
they allow us to do certain integrals in closed form; they also 
give the physical result that the particle production vanishes in 
the limit that the solenoid flux goes to zero. The values of $k$ 
now take on discrete values for finite $a$, but the set becomes 
a continuum in the limit $a \to \infty$. Then the singular   
gauge eigenmodes are 
\begin{equation}
\psi_{k q l} (r, \theta, z) = 
\sqrt{\frac{k}{2La}} ~
J_{| l - \alpha |} (k r) e^{i q z} e^{i (l - \alpha ) \theta }
\label{eq:modes}
\end{equation}
where $ 0 \leq \theta < 2 \pi$, and $\alpha = \epsilon /2\pi$. 
The modes are labelled by the radial momentum $k$, the $z$-momentum, 
$q$, and the azimuthal quantum number $l$. The frequency of the mode is
\begin{equation}
\omega_{kq} = \sqrt{ k^2 + q^2 + m^2}
\label{eq:modefreq}
\end{equation}
Note that the eigenmodes have a discontinuity along the positive 
$x$-axis. This is because we are working in singular gauge where  
the gauge potential is singular on the positive $x-$axis. 

The normalization integral of the modes is
\begin{equation}
\int_0^a d r \hspace{1mm}r  \int_0^{2 \pi} d \theta \int_0^L d z ~
\psi_{k' q' l'}^{\ast}  \psi_{k q l} = 
\delta_{l l'} \delta_{q q'} \delta_{k k'}.
\label{eq:normalization}
\end{equation}
where we have taken $ka \gg | (l-\alpha)^2-1/4|$ and used the 
asymptotic form of the Bessel functions
\begin{equation}
J_\nu (x) \approx \sqrt{\frac{2}{\pi x}} \cos \left (
       x - \frac{\nu\pi}{2} - \frac{\pi}{4} \right )
\end{equation}
together with the integral
\begin{eqnarray}
\int_0^a dr ~ r ~ J_\nu (kr) J_\nu (k'r) =&& \nonumber \\ 
&&
\hskip -3 cm \frac{a^2}{2} [ (J_\nu (ka) )^2 - J_{\nu-1}(ka)J_{\nu+1}(ka) ]
\delta_{kk'}
\end{eqnarray}

As in the previous section we assume that the solenoid oscillates 
in the $xz-$plane while remaining aligned with the $z$-axis. The 
modes for a shifted solenoid are obtained by appropriately translating 
the modes eq (\ref{eq:modes}). Hence the transition element
designated $a_{\beta \alpha}$ in the appendix, Eqs.~(\ref{eq:aandw}) 
and (\ref{eq:oscillatorytransition}), is here given by 
\begin{equation}
a_{k'q'l', kql} = \int_0^a dr \hspace{1mm} r \int_0^{2 \pi} d \theta 
\int_0^L d z ~ \psi^{\ast}_{k'q'l'} \frac{\partial}{\partial x} \psi_{kql}. 
\label{eq:transitionx}
\end{equation}
A straightforward calculation reveals 
\begin{equation}
a_{k' q' l', k q l}  =  
  - L \delta_{q q'} \frac{2 k \sin(\pi \alpha)}{k'^2 - k^2} 
                       \left( \frac{k'}{k} \right)^{\alpha} 
\label{eq:transitionone}
\end{equation}
for $l = 0$, $l' = 1$ and 
\begin{equation}
a_{k' q' l', k q l} = 
  - L \delta_{q q'} \frac{2 k \sin(\pi \alpha) }{k'^2 - k^2} 
                   \left( \frac{k'}{k} \right)^{\alpha}
\label{eq:transitiontwo}
\end{equation}
for $l = 1$, $l' = 0$; for all other combinations of $l$ and $l'$ it vanishes. 

Using Eq.~(\ref{eq:rate}) and taking the continuum limit 
$L, a \rightarrow \infty$ we obtain 
$d {\dot N}' = d k ~ d q ~ {\cal I} (k, q)$ where
\begin{eqnarray}
{\cal I} (k, q) & = & \frac{v_0^2 \sin^2 (\pi\alpha)}{16\pi^2} 
\frac{k~ k_c^2}{\omega_{kq} \Omega^2} 
\left[ 
     \left ( \frac{k}{k_c} \right )^{2\alpha}
 + \left( \frac{k_c}{k} \right)^{2 \alpha -2} \right] 
\nonumber \\
 && \times 
\Theta( \Omega - \omega_{kq} - \sqrt{ q^2 + m^2 } )
\label{eq:ikq}
\end{eqnarray}
is the rate of particle production at wave-vector $(k, q)$. 
(To include antiparticles, we would double this rate.)
Here $k_c \equiv ( \Omega^2 + k^2 - 2 \Omega \omega_{kq} )^{1/2}$. 
Eq.~(\ref{eq:ikq}) applies for $ 0 < \alpha < 1$; by virtue of 
the periodicity of the scalar field modes,
$I(k,q)$ is a periodic function of $\alpha $ with period 1. 

Some of the quantities in Eq.~(\ref{eq:ikq}) have a straightforward 
interpretation in terms of the kinematics of pair production. Let 
the pair of particles have momenta $(k, q)$ and $(k', q')$. By 
conservation of $z$-axis momentum $q + q' = 0$. By energy conservation 
$\omega_{kq} = \Omega - \omega_{k'q'}$. The minimum value of 
$\omega_{k'q'}$ is $\sqrt{ q^2 + m^2 }$; hence 
$\omega_{kq} < \Omega - \sqrt{ q^2 + m^2 }$; 
this accounts for the $\Theta$ function in Eq.~(\ref{eq:ikq}). 
Similarly if one particle has momentum $(k,q)$, then $k_c$ 
represents the radial momentum of the second particle. 

In the limit $\alpha \rightarrow 0$ Eq.~(\ref{eq:ikq}) 
simplifies to 
\begin{equation}
{\cal I} (k, q) = \frac{\alpha^2}{16} \frac{v_0^2}{\Omega^2} 
                 \frac{k}{\omega_{kq}} ( k_c^2 + k^2 ) 
                 \Theta( \Omega - \omega_{kq} - \sqrt{ m^2 + q^2 } ).
\label{eq:smallflux}
\end{equation}
To make contact with the results of the previous subsection we 
consider the non-relativistic limit of the result for $d{\dot N}'$, 
(Eq.~(\ref{ddotNprime})) 
in which only the first harmonic contribution is significant.   
If we write ${\bm p}$ in cylindrical co-ordinates as 
${\bm p} = (k, \theta, q)$, integrate $d{\dot N}'$ over ${\bm p}'$ 
and $\theta$, halve the result since the relativistic expression
counts production of particles and antiparticles, we obtain 
the flux in Eq.~(\ref{eq:smallflux}). Thus the non-relativistic 
limit of the small flux approximation coincides with the small flux 
limit of the non-relativistic moving solenoid approximation, providing 
a useful check on both calculations.

\section{AB Radiation from Cosmic String Loops}
\label{loopsAB}

A cosmic string loop trajectory is written in terms of 
left- and right- movers
\begin{equation}
{\bm X} = \frac{1}{2} [ {\bm a}(\sigma - \tau) +
                        {\bm b}(\sigma + \tau) ]
\label{loopX}
\end{equation}
The choice of gauge, Eq.~(\ref{gaugeconditions}), requires
\begin{equation}
|{\bm a}'| = 1 = |{\bm b}'|
\end{equation}
where primes denote derivatives with respect to the argument. 

Insertion of Eq.~(\ref{loopX}) in (\ref{amplitude}) with some
simplifications gives
\begin{equation}
i{\cal M} = \frac{\epsilon}{4} ~\epsilon_{\mu\nu\alpha\beta} 
            \frac{{(p+p')^\mu} (p-p')^\nu}{(p+p')^2} I_+^\alpha I_-^\beta
\end{equation}
where
\begin{eqnarray}
I_+^\alpha &=& \int_{-\infty}^{+\infty} d\sigma_+ {b^\alpha}'
                    e^{-ik\cdot b/2}
\nonumber \\
I_-^\alpha &=& \int_{-\infty}^{+\infty} d\sigma_- {a^\alpha}'
                    e^{-ik\cdot a/2}
\label{Ipmdefn}
\end{eqnarray}
and $\sigma_\pm \equiv \sigma \pm \tau$ and $k=p+p'$. Note that the
integrations have been extended to the entire $\sigma,t$ plane. This
is valid since the integrands are periodic functions in both $\sigma$
and $t$, and the only effect of the larger domain of integration will 
be to yield Dirac delta functions instead of Kronecker delta's.

The integrals, $I_\pm$ satisfy the orthogonality relations
\begin{equation}
k \cdot I_\pm = 0 
\label{orthogonality}
\end{equation}
which allow us to relate $I_\pm^0$ to ${\bm I}_\pm$,
\begin{equation}
k^0 I_\pm^0 = {\bm k}\cdot {\bm I}_\pm \ .
\end{equation}
By using these relations, we find
\begin{eqnarray}
S &\equiv& \epsilon_{\mu\nu\alpha\beta} ~ {p^\mu}' p^\nu
                 I_+^\alpha I_-^\beta         \nonumber \\
&=& \frac{k_\mu k^\mu}{2k^0} ( {\bm p}-{\bm p}')
             \cdot ({\bm I}_+ \times {\bm I}_-)
\label{Smne0}
\end{eqnarray}

Now we can write
\begin{eqnarray}
|{\cal M}|^2 &=& \frac{\epsilon^2}{4(k_\mu k^\mu)^2} |S|^2 \nonumber \\
             &=& \frac{\epsilon^2}{16(\omega +\omega ')^2} 
     |({\bm p}-{\bm p}') \cdot {\bm I}_+ \times {\bm I}_- |^2
\end{eqnarray}

The integrations in Eqs.~(\ref{Ipmdefn}) extend over an infinite
range whereas the loop dynamics is periodic. We can use the
series representation of the Dirac delta function
\begin{equation}
\delta (x) = \frac{1}{2\pi} \sum_{n=-\infty}^{+\infty} e^{inx} \ , 
                 \ \ -\pi < x < +\pi
\end{equation}
to write the integrals over only one periodic domain
\begin{eqnarray}
I_+^\alpha &=& \sum_{n=1}^\infty \int_0^{L} d\sigma_+ {b^\alpha}'
                    e^{-ik\cdot b/2} 
    \delta \left ( \frac{k^0 L}{4\pi} -n \right )
\nonumber \\
I_-^\alpha &=& \sum_{n=1}^\infty \int_0^L d\sigma_- {a^\alpha}'
                    e^{-ik\cdot a/2}
    \delta \left ( \frac{k^0 L}{4\pi} -n \right )
\label{Ipmseries}
\end{eqnarray}
The integrals ${\bm I}_\pm$ are non-vanishing only for discrete 
positive values of the total energy 
\begin{equation}
k^0 = \omega + \omega ' = \frac{4\pi n}{L}
\label{sump0}
\end{equation}
where $n$ is a positive integer. Therefore it is convenient
to write the squared amplitude for a fixed harmonic
\begin{equation}
|{\cal M}|_n^2 = \left ( \frac{\epsilon L}{16\pi n } \right )^2
    |({\bm p}-{\bm p}') \cdot {\bm I}_+^{(n)} \times {\bm I}_-^{(n)} |^2
\end{equation}
where ${\bm I}_\pm^{(n)}$ refer to the integrals in Eq.~(\ref{Ipmseries}) 
together with the constraint in Eq.~(\ref{sump0}).

The squared amplitude can now be written as
\begin{equation}
|{\cal M}|^2 = \frac{\pi}{L} \sum_{n=1}^\infty
    |{\cal M}|_n^2 
    \delta \left ( \frac{k^0 L}{4\pi} -n \right )
    \delta_E (0)
\end{equation}
where $\delta_E (0)$ is an energy conservation delta function
and we will divide out by $2\pi \delta_E (0)$ to obtain particle 
radiation per unit time.

We will now find the radiation in the $m=0$ case
from two different kinds of loops, 
one that contains kinks (sharp corners where the tangent vector is 
discontinuous) and the other that contains cusps (points on the 
loop that momentarily reach the speed of light).

\subsection{Kinky Loop}
\label{kinkyloop}

Here we consider the ``degenerate kinky loop''
\begin{eqnarray}
{\bm a} &=& \sigma_- {\bm A} , \ \ \ 
            0 \le \sigma_- \le L/2 \nonumber \\
{\bm a} &=& (L - \sigma_- ) {\bm A} , \ \ \ 
            L/2 \le \sigma_- \le L  \nonumber \\
{\bm b} &=& \sigma_+  {\bm B} , \ \ \ 
            0 \le \sigma_+ \le L/2 \nonumber \\
{\bm b} &=& (L - \sigma_+ ) {\bm B} , \ \ \ 
            L/2 \le \sigma_+ \le L 
\end{eqnarray}
where ${\bm A}$ and ${\bm B}$ are two fixed unit vectors.
Also note that
\begin{equation}
a^0 = - \sigma_- \ , \ \ \ b^0 = \sigma_+
\end{equation}
so that $X^0=t$. 

\begin{figure}
\scalebox{1.0}{\includegraphics[angle=0]{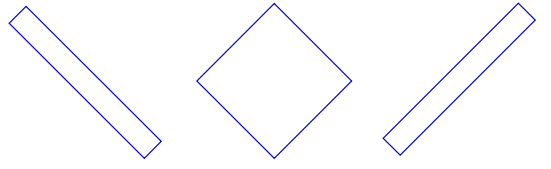}}
\caption{Three snapshots of a degenerate kinky loop.}
\label{degkinkyloop}
\end{figure}

This loop is ``degenerate'' because it consists of four straight
segments and is ``kinky'' because of its four corners. The four
straight segments propagate with constant speed but shrink and 
expand due to the motion of the kinks (see Fig.~\ref{degkinkyloop}). 

Now the relevant integrals can be evaluated in the $m=0$ case and
with $k^0=4\pi n/L$, 
\begin{eqnarray}
\hskip -0.5 cm
{\bm I}_-^{(n)} &=& \frac{2L}{n\pi} 
    \frac{e^{i(1+{\bm \kappa}\cdot {\bm A})n\pi /2}}
                {1-({\bm \kappa}\cdot {\bm A})^2}
       \sin \left ( (1+{\bm \kappa}\cdot {\bm A}) \frac{n\pi}{2} \right )
       {\bm A} \\
\hskip -0.5 cm
{\bm I}_+ ^{(n)} &=& \hskip -0.2 cm 
\frac{2L}{n\pi} \frac{e^{-i(1-{\bm \kappa}\cdot {\bm B})n\pi /2}}
                {1-({\bm \kappa}\cdot {\bm B})^2}
       \sin \left ( (1-{\bm \kappa}\cdot {\bm B}) \frac{n\pi}{2} \right )
       {\bm B}
\label{Ipmn}
\end{eqnarray}
where ${\bm \kappa} \equiv {\bm k}/k^0 = L({\bm p}+{\bm p}')/4\pi n$. 
This gives
\begin{eqnarray}
|{\cal M}|_{m=0; n}^2 &=& \frac{\epsilon^2}{16}
      \left ( \frac{L}{n\pi} \right )^6
              |({\bm p}-{\bm p}') \cdot {\bm A} \times {\bm B} |^2
   \nonumber \\
&&
\hskip -1.5 cm 
\times
\frac{\sin ^2\left ( (1+{\bm \kappa}\cdot {\bm A}) \frac{n\pi}{2} \right )
      \sin ^2\left ( (1-{\bm \kappa}\cdot {\bm B}) \frac{n\pi}{2} \right )}
     { [{1-({\bm \kappa}\cdot {\bm A})^2}]^2
                [{1-({\bm \kappa}\cdot {\bm B})^2}]^2 }
\label{calMn}
\end{eqnarray}

The expression in (\ref{calMn}) is finite as can be seen by taking 
limits, say as ${\bm \kappa}\cdot {\bm A} \to 1$. This shows
that kinks on loops give off finite AB radiation at every harmonic. 
(The sum over harmonics, however, will be seen to diverge.) Maximum 
emission occurs when ${\bm p}-{\bm p}'$ is parallel to 
${\bm A} \times {\bm B}$ 
{\it i.e.} perpendicular to the plane of the loop. Also, the factors 
depending on ${\bm \kappa}$ are maximum for     
${\bm \kappa}\cdot {\bm A}=0={\bm \kappa}\cdot {\bm B}$, implying that 
${\bm \kappa}$, hence ${\bm p}+{\bm p}'$, is perpendicular to the 
plane of the loop. Since the amplitude vanishes for ${\bm p}-{\bm p}'=0$,
we conclude that the maximum emission occurs for
${\bm p} = -{\bm p}'$ and perpendicular to the plane of the loop.

The differential rate of particle production is given by 
Eq.~(\ref{eq:dN})
\begin{eqnarray}
d{\dot N} &=& \frac{1}{2L} \frac{d^3p}{(2\pi)^3} \frac{1}{2\omega} 
     \frac{d^3p'}{(2\pi)^3} \frac{1}{2\omega'}  \nonumber \\
          &\times& \sum_{n=1}^\infty |{\cal M}|_{m=0; n}^2 
          \delta(n-(\omega + \omega ')L/4\pi  ) 
\end{eqnarray}

The total energy emitted is
\begin{equation}
{\dot E} = \sum_{n=1}^\infty {\dot E}_n
\end{equation}
where
\begin{eqnarray}
{\dot E}_n &=& \frac{2\pi}{L^2}  n
 \int \frac{d^3p}{(2\pi)^3} \frac{1}{2\omega} 
 \int \frac{d^3p'}{(2\pi)^3} \frac{1}{2\omega'}  \nonumber \\
          &\times& |{\cal M}|_{m=0; n}^2 
           \delta(n-(\omega + \omega ')L/4\pi  ) 
\label{powern}
\end{eqnarray}
The integral can be evaluated by numerical means. 

\begin{figure}
\scalebox{0.33}{\includegraphics[angle=-90]{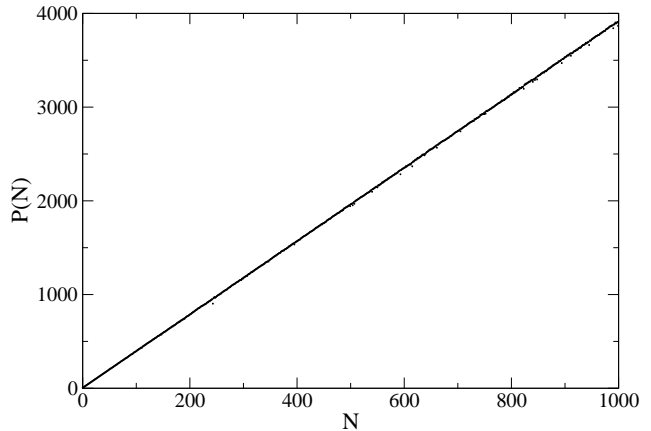}}
\caption{The plot of $2\pi (\pi^2 L/4\epsilon)^2P (N)$ versus $N$ 
for the degenerate kinky loop shows $P(N) \propto N$ growth. 
}
\label{kinkydotEnvsn}
\end{figure}

However it is prudent to check the scaling with $n$ before 
performing detailed calculation.
Due to the energy conservation delta function we estimate 
$\omega \sim n$, $\omega '\sim n$, $d^3p \sim n^3$,
$d^3p' \sim n^3$, and the delta function contributes
$n^{-1}$. The contribution of the factor $|{\cal M}|_{m=0;n}^2$ 
is less clear: a factor of $n^{-4}$ is left over after re-scaling 
all the momenta and there may also be factors of $n$ due to 
contributions from the trignometric functions
as suggested by considering the similar integral
\begin{eqnarray}
I &=& \int_0^1 dx \frac{\sin^2(nx)}{x^2} 
 = n \int_0^n dy \frac{\sin^2(y)}{y^2} \nonumber \\
  &\to& n \int_0^\infty dy \frac{\sin^2(y)}{y^2} = O(n^{+1})
\end{eqnarray}
To find the scaling, we have summed ${\dot E}_n$ 
over $n$ up to a cutoff $N$ using
\begin{eqnarray}
\sum_{n=1}^N \sin^2(nx) \sin^2(ny) &=&
\frac{M}{8} - \frac{\sin (Mx)}{8\sin x} - \frac{\sin (My)}{8\sin y} 
\nonumber \\
&+&\frac{\sin(Mx_-)}{16 \sin x_-} + \frac{\sin(Mx_+)}{16 \sin x_+}  
\end{eqnarray}
where $M =2N+1$ and $x_\pm = x\pm y$. Then the remaining
integrations are done numerically on Mathematica to evaluate
$$
P(N) = \sum_{n=1}^N {\dot E}_n
$$ 
In Fig.~\ref{kinkydotEnvsn}, we
plot $P(N)$ versus $N$ for the loop with ${\bf A}={\hat x}$ and 
${\bf B}={\hat y}$, clearly showing that the total power radiated
grows lineary with the upper cutoff $N$.

Since $P(N) \propto N$, the total power emitted diverges linearly
and
\begin{equation}
P(N) \sim \frac{\epsilon^2}{L^2} N
\end{equation}
The kink width provides a cutoff on harmonics and we assume
this is similar to the width of the string, $1/\sqrt{\mu}$. 
Therefore it is reasonable to take $N \sim \sqrt{\mu} L$ and 
estimate the total power emitted,
\begin{equation}
{\dot E} \sim \epsilon^2 \mu \left ( \frac{1}{\sqrt{\mu}L} \right ) 
\label{dotEcuspy}
\end{equation}
This energy emission rate is much larger than the naive
dimensional analysis result $\epsilon^2/L^2$, by a factor of
$\sqrt{\mu}L \sim L/w$ where $w$ is the width of the string.

\subsection{Cuspy Loop}
\label{cuspyloop}

We now consider smooth loops that have cusps {\it i.e.} points 
that reach the speed of light.

A simple loop trajectory is given by
\begin{eqnarray}
a^0 &=& -\sigma_- , \ \ 
{\bm a} = \frac{L}{2\pi} (\sin s_-, -\cos s_-, 0) \nonumber \\
b^0 &=& \sigma_+ , \ \ 
{\bm b} = \frac{L}{2\pi} (\sin s_+, 0, -\cos s_+) 
\label{loopab}
\end{eqnarray}
where $s_\pm = 2\pi \sigma_\pm /L$.
The loop has cusps because there are points such that
${\bm a}'= - {\bm b}'$. These occur at $s_+ =s_- = 0, \pi$,
at which point the velocity of the string is 
${\bm v} = \pm {\hat x}$ {\it i.e.} the point on the string 
reaches the speed of light.

Inserting the loop trajectory in Eq.~(\ref{Ipmseries}) gives
\begin{eqnarray}
{\bm I}_-^{(n)} &=&  (-1)^n e^{in \phi_-} L 
\left [ \frac{J_n(n \kappa_{xy})}{\kappa_{xy}} {\hat {\bm \kappa}}_{xy}
     -i J_n' (n \kappa_{xy}) {\hat {\bm \kappa}}^\perp_{xy} \right ] 
\nonumber \\
{\bm I}_+^{(n)} &=& e^{-in\phi_+} L 
\left [ \frac{J_n(n \kappa_{xz})}{\kappa_{xz}} {\hat {\bm \kappa}}_{xz}
     -i J_n' (n \kappa_{xz}) {\hat {\bm \kappa}}^\perp_{xz} \right ] 
\end{eqnarray}
where $\kappa^\mu = L k^\mu /(4\pi n)$, 
\begin{eqnarray}
{\bm \kappa}_{xz} &\equiv& (\kappa_x,0,\kappa_z) \ , \ \ \ 
{\bm \kappa}_{xy} \equiv (\kappa_x,\kappa_y,0) \nonumber \\
{\bm \kappa}^\perp_{xz} &\equiv& (-\kappa_z,0,\kappa_x) \ , \ \ \ 
{\bm \kappa}^\perp_{xy} \equiv (\kappa_y,-\kappa_x,0) 
\end{eqnarray}
and we have defined the unit vectors ${\hat \kappa}_{xz}$,
${\hat \kappa}_{xy}$, and
\begin{equation}
\tan \phi_+ = \frac{\kappa_z}{\kappa_x} \ , \ \ \ 
\tan \phi_- = \frac{\kappa_y}{\kappa_x} \ . 
\end{equation}

The rate of energy emission is obtained from Eq.~(\ref{powern}).
We rescale all momenta and energy by $L/4\pi n$ and obtain
\begin{eqnarray}
{\dot E}_n &=& \frac{\epsilon^2 n^4}{2\pi L^2}  
  \int_0^1 dq \int d^2{\hat p} \int d^2 {\hat p}' ~ q(1-q)
   \times
\nonumber \\
&& 
\hskip -0.5 cm
\biggl | (q{\hat p} - (1-q){\hat p}' ) \cdot
 \biggl \{ \frac{J_n(n \kappa_{xy})}{\kappa_{xy}} {\hat \kappa}_{xy}
 - i J_n' (n \kappa_{xy}) {\hat \kappa}_{xy}^\perp \biggr \} 
\nonumber \\
&& \hskip 1 cm \times
    \biggl \{ \frac{J_n(n \kappa_{xz})}{\kappa_{xz}} {\hat \kappa}_{xz}
             - i J_n' (n \kappa_{xz}) {\hat \kappa}_{xz}^\perp \biggr \}
   \biggr |^2
\label{cuspdotEnexp}
\end{eqnarray}
where ${\bm \kappa} = q{\hat p} + (1-q){\hat p}'$.

Note that the integrand is non-singular because of the properties
of the Bessel functions (see Fig.~\ref{besseloverx}). 
The integration is over a compact region and can be evaluated 
numerically. For the evaluation it is advantageous to rewrite 
Eq.~(\ref{cuspdotEnexp}) using the Bessel function identities
\begin{equation}
2 n \frac{J_n(x)}{x} = J_{n-1}(x) + J_{n+1}(x)
\end{equation}
\begin{equation}
2 J_n'(x) = J_{n-1}(x)-J_{n+1}(x)
\end{equation}

To estimate the dependence of ${\dot E}_n$ on $n$, we compare
the present integration to that for the kinky loop. The main 
difference is that the integrand for the kinky loop contains 
trignometric functions whose argument contains $n$, whereas 
here we have Bessel functions.
As larger values of $n$ are considered, the Bessel functions
contribute for smaller windows of $\kappa_{xy}$ and $\kappa_{xz}$,
whereas the periodic trignometric functions contribute for smaller
windows but there are more such windows because of the periodic
nature of the trignometric functions. Following the argument below 
Eq.~(\ref{powern}), we expect ${\dot E}_n \propto n^0$ and
hence $P(N) \propto N$ for the cuspy loops.

\begin{figure}  
\scalebox{0.33}{\includegraphics[angle=-90]{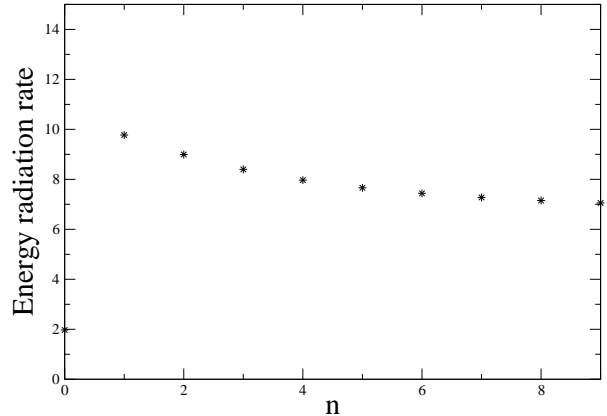}}
\caption{$(8\pi L^2/\epsilon^2){\dot E}_n$ versus $n$ for the loop
with cusps.} 
\label{cuspydotEnvsn}
\end{figure}

We have numerically evaluated the integration in 
Eq.~(\ref{cuspdotEnexp}) and, in Fig.~\ref{cuspydotEnvsn}, we 
plot $8\pi L^2 {\dot E}_n/\epsilon^2$ versus $n = 1,\ldots, 10$. The 
plot confirms the ${\dot E}_n \propto n^0$ estimate for the cuspy 
loops.

The string thickness, $w \sim 1/\sqrt{\mu}$ provides a 
natural cutoff on the wavelength of the emitted radiation.
Thus the maximum harmonic $\sim L/w$ and we estimate the 
total AB power for massless particles from the cuspy loop 
\begin{equation}
{\dot E} \sim \frac{\epsilon^2}{L^2} N 
         \sim \frac{\epsilon^2}{L^2} \frac{L}{w}
\label{cuspydotE}
\end{equation}
This is of the same order of magnitude as the radiation from kinks.

The emission of massive particles is conceptually similar but 
technically more involved. We have left that analysis for
future work.

\section{Gravitational AB Radiation}
\label{gravAB}

\begin{figure}
\scalebox{1.0}{\includegraphics{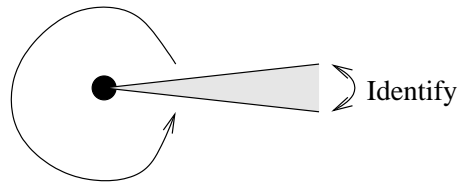}}
\caption{A conical space is obtained by cutting out a
wedge from a Euclidean two dimensional space and identifying
the edges. The wavefunction of a particle that goes around
the tip of the cone, acquires a phase given by the deficit
angle.
}
\label{conicalmetric}
\end{figure}

The above analysis for AB radiation suggests that a corresponding
effect must exist in a gravitational setting. Consider the conical
metric of a straight cosmic string (see Fig.~\ref{conicalmetric}).
When {\it any} particle goes around the string and returns to its 
original position, the wavefunction acquires a phase factor given 
by the conical deficit angle $\delta = 8\pi G\mu$, where $\mu$
is the string tension. Hence any particle will interact with a
cosmic string by the gravitational AB effect and we expect
an oscillating cosmic string to radiate gravitational AB radiation
that contains {\it all} species of particles. 

The existence of gravitational AB radiation may also be argued
on the grounds that an oscillating cosmic string provides a time 
dependent background for quantum fields. All the vacuum modes
of the quantum fields have to adjust to the dynamical background.
Then it is natural to expect some ``sloshing'' {\it i.e.} pair 
creation. The process has been studied in \cite{Garriga:1989bx}
and here we extend that analysis to the case of kinks. 

The most intriguing aspect of gravitational AB radiation is that 
it implies that cosmic strings will radiate photons. Based on 
our calculations of the previous sections, we expect the emission
to be largest from kinks on loops.
We will now sketch a calculation of the gravitational AB
radiation of massless scalars from a degenerate, kinky loop. 
The relevant interaction between the metric and the 
scalar field can be obtained if we start with the action for 
a massless scalar field
\begin{equation}
S_\phi = \int d^4x \sqrt{-g} ~ \frac{1}{2} 
         g^{\mu\nu} \partial_\mu\phi \partial_\nu\phi
\end{equation}
and expand around the Minkowski metric, 
\begin{equation}
g_{\mu\nu}=\eta_{\mu\nu}+h_{\mu\nu}\ , \ \ 
g^{\mu\nu}=\eta^{\mu\nu}-h^{\mu\nu}
\end{equation}
This gives
\begin{equation}
S_\phi = S_{\phi,0} - \int d^4x \frac{1}{2} 
         H^{\mu\nu} \partial_\mu\phi \partial_\nu\phi + \ldots
\end{equation}
where $S_{\phi,0}$ is the action in a Minkowski background, and
\begin{equation}
H_{\mu\nu} = 
      h_{\mu\nu} - \frac{1}{2} \eta_{\mu\nu} h^\alpha_\alpha
\end{equation}

The metric perturbation, $h_{\mu\nu}$, is sourced by the string
and can be written in Fourier space as
\begin{equation}
{\tilde h}_{\mu\nu} = -\frac{16\pi G}{k^2} 
 \left ( {\tilde T}^{(s)}_{\mu\nu} -
         \frac{1}{2} \eta_{\mu\nu} {\tilde T}^{{(s)}\alpha}_\alpha 
 \right )
\end{equation}
where the superscript $(s)$ on ${\tilde T}$ denotes that this is the 
Fourier transform of the string energy-momentum tensor which is
given by \cite{VilenkinShellard}
\begin{equation}
T^{(s)}_{\mu\nu}(x) = \mu \int d^2\sigma 
    ({\dot X}_\mu {\dot X}_\nu
      - X'_\mu X'_\nu) \delta^{4}(x-X) 
\end{equation}
Now the string energy-momentum tensor can be written in terms of
$a(\sigma_-)$ and $b(\sigma_+)$.
\begin{equation}
T^{(s) \mu\nu} = -\frac{\mu}{4} \int d^2\sigma_\pm 
      ({a^\mu}' {b^\nu}' + {a^\nu}' {b^\mu}' ) \delta^4(x-X)
\end{equation}
The Fourier transform can be done and it is clear that 
it will factorize into integrations over left-moving 
($\sigma_-$) and right-moving ($\sigma_+$) variables.
For our present analysis, we would like to focus
on the degenerate, kinky loops, for which ${a^\mu}'$
and ${b^\mu}'$ are piecewise continuous. Then
\begin{equation}
{\tilde T}^{(sq)\mu\nu} = - \frac{\mu}{4} 
                       (I_+^\mu I_-^\nu + I_+^\nu I_-^\mu )
\end{equation}
where $I_\pm^\mu$ have the same meaning as in Eq.~(\ref{Ipmdefn}).

In momentum space we have
\begin{equation}
{\tilde H}_{\mu\nu} = -\frac{16\pi G}{k^2} {\tilde T}_{\mu\nu} 
\end{equation}
and the pair production amplitude from the square loop is
\begin{eqnarray}
i{\cal M}^{(sq)} &=&
-\frac{8\pi G}{k^2} 
 {\tilde T}^{(sq)}_{\mu\nu} (p^\mu {p^\nu}'+p^\nu {p^\mu}') 
    \nonumber \\
&=& \frac{\delta}{8p\cdot p'} ( p\cdot I_+ ~ p'\cdot I_-
         + p\cdot I_- ~ p'\cdot I_+ )
\end{eqnarray}
We use the orthogonality relations in Eqs.~(\ref{orthogonality}),
together with energy conservation, to simplify the amplitude 
expression as in Sec.~\ref{kinkyloop}, to eventually obtain
\begin{eqnarray}
|{\cal M}^{(sq)}|^2 &=& 4~ \delta^2 \left ( \frac{L}{n\pi} \right )^4
 q^2(1-q)^2 [{\hat q}\cdot {\bm A} ~ {\hat q} \cdot {\bm B} ]^2 \nonumber \\
&& 
\hskip -1.5 cm 
\times
\frac{\sin ^2\left ( (1+{\bm \kappa}\cdot {\bm A}) \frac{n\pi}{2} \right )
      \sin ^2\left ( (1-{\bm \kappa}\cdot {\bm B}) \frac{n\pi}{2} \right )}
     { [{1-({\bm \kappa}\cdot {\bm A})^2}]^2
                [{1-({\bm \kappa}\cdot {\bm B})^2}]^2 }
\label{calM2square}
\end{eqnarray}
where, as in Sec.~\ref{kinkyloop},
\begin{equation}
{\bm \kappa} = \frac{L({\bm p}+{\bm p}')}{4\pi n}
\end{equation}
and
\begin{equation}
q = \frac{L|{\bm p}|}{4\pi n} \ , \ \ 
{\hat q} = \frac{{\hat p}'-{\hat p}}{|{\hat p}'-{\hat p}|}
\end{equation}

\begin{figure}
\scalebox{0.33}{\includegraphics[angle=-90]{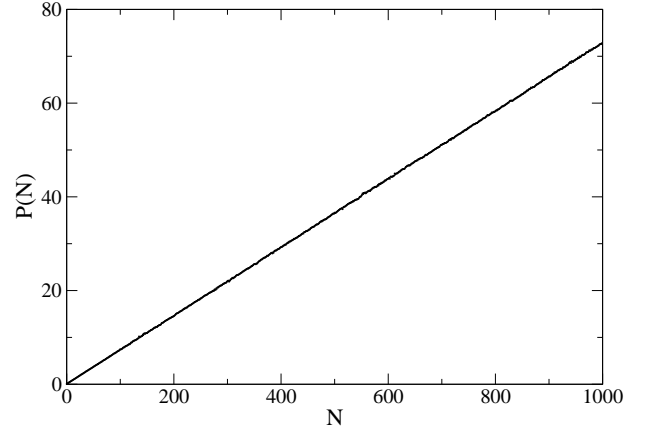}}
\caption{Plot of the power radiated, $2\pi (\pi^2 L/8\delta)^2 P(N)$ up
to some harmonic $N$ for gravitational AB radiation from the square 
loop, showing the $P(N) \propto N$ growth. 
}
\label{gravABplot}
\end{figure}

The expression in Eq.~(\ref{calM2square}) for gravitational AB 
radiation from square loops is very similar to the expression for
gauge AB radiation from degenerate, kinky loops, given in 
Eq.~(\ref{calMn}). The main difference is in the terms preceding
the trignometric functions. The 
$[{\hat q}\cdot {\bm A} ~ {\hat q} \cdot {\bm B} ]^2$ factor
implies that gravitational AB radiation is in the plane of
the loop, in contrast to gauge AB radiation. However, the 
scaling with $n$ follows the discussion below Eq.~(\ref{powern}), 
and is expected to be independent of the form of the prefactor. 
As a check on the argument, we have evaluated $P(N)$ 
numerically and the results are shown in Fig.~\ref{gravABplot},
confirming that $P(N)\propto N$ {\it i.e.} 
${\dot E}_n \propto n^{0}$. 

So far we have outlined gravitational AB radiation into massless
scalars. We can proceed similarly for the radiation of photons.
Now the linear order interaction is
\begin{equation}
S = \frac{1}{2} \int d^4x 
\left ( h^{\mu\nu} - \frac{1}{4} \eta^{\mu\nu} h^\alpha_\alpha \right )
      F_{\mu\beta}F_\nu^{~ \beta}
\end{equation}
where the field strength is defined in terms of the gauge potential
as $F_{\mu\nu} = \partial_\mu A_\nu - \partial_\nu A_\mu$ and indices
are raised and lowered by the Minkowski metric. The rest of the
calculation is expected to be very similar to that for massless
scalars, though the final state will also depend on photon polarizations.
We expect that there will be string-like signatures contained 
in the polarizations, ${\bm \epsilon}$ of the emitted photons, 
{\it e.g.} linear polarization along the string due to factors like
${\bm \epsilon} \cdot {\bm A}$ in the amplitude, and these 
would be worth exploring further.

As in the gauge AB radiation case, the $P(N) \propto N$
({\it i.e.} ${\dot E}_n \propto n^0$)
result suggests that the power radiated in {\it photons} due 
to the gravitational AB effect can be estimated as
\begin{equation}
{\dot E}_{AB~\gamma} \sim \delta^2 ~ \mu 
         \left ( \frac{1}{\sqrt{\mu} L} \right )
\end{equation}
The loop also emits power in gravitational waves, 
\begin{equation}
{\dot E}_{gw} \sim \delta ~ \mu 
\end{equation}
Hence
\begin{equation}
{\dot E}_{AB~\gamma} \sim  {\dot E}_{gw} ~ \delta ~
                           \left ( \frac{w}{L} \right ) 
\end{equation}
where $w$ denotes the width of the string.
Even though the power emitted in photons is a very small fraction 
of the power emitted in gravitational waves, it can still be 
of interest, especially because it is in photons. For 
$G\mu = 10^{-8}$, we can write $\mu \sim 10^{51} ~ {\rm ergs/s}$, 
from which we get ${\dot E}_g \sim 10^{43} ~ {\rm ergs/s}$ and 
${\dot E}_\gamma \sim 10^{35} (w/L) ~ {\rm ergs/s}$. 
The lifetime of the loop is determined by its gravitational
radiation rate and is estimated as $\sim L/\delta$. Hence
the total energy emitted in photons is $\sim \delta \sqrt{\mu}$.
With $\delta \sim 10^{-8}$, the total energy emitted in
photons $\sim 10^7$ GeV, and is independent of the size
of the loop.

The number of photons emitted per unit time in the $n^{\rm th}$
harmonic, ${\dot N}_n$, is proportional to $n^{-1}$. Hence 
the same number of photons are emitted in every logarithmic frequency
interval
\begin{equation}
d{\dot N}_\omega \sim \frac{\delta^2}{L} \frac{d\omega}{\omega} 
\end{equation}
 
In \cite{Garriga:1989bx} the accummulated photon background due 
to {\it cusps} on a network of cosmic strings was considered. 
Here we estimate photon emission from kinks along a single long 
string in the present universe. This is relevant for direct detection
of a string. The signal may be stronger than that from cusps
particularly because kinks are expected to be much more numerous 
on strings as compared to cusps. A recent estimate of the number 
density of kinks on a string is \cite{Copeland:2009dk}
\begin{equation}
n_k \sim \frac{1}{t} \left ( \frac{t}{t_*} \right )^{0.9}
\end{equation}
where $t_*$ is the epoch at which friction on the string network 
dynamics ceases to be important compared to the Hubble damping. 
Based on the scattering rate of particles off the string, a rough 
estimate is $t_* \sim t_f/\delta$ where $t_f$ is the time of 
formation of the strings.

To obtain numerical estimates, we have the present cosmological 
epoch at $t_0 \sim 10^{17}$ s, and for strings with $\delta =10^{-8}$, 
we find $t_* \sim 10^{-27}$ s. Thus $t_0/t_* \sim 10^{44}$. 
With each kink emitting a total energy of $\delta \sqrt{\mu}$ over
a Hubble time, the emission per unit length of string per unit time 
is $\sim \delta \sqrt{\mu} (t_0/t_*)/t_0^2 \sim 10^4$ ergs/cm-s,
which is $\sim 10^{25}$ ergs/kpc-s. The energy emitted from 
strings on astrophysical length scales is quite small compared 
to astrophysical sources (e.g. the solar luminosity is $\sim
10^{33}$ ergs/s), though the unique stringy features, (such
as linear emission, polarization, flat spectrum) of the
emission may help observations. Even if the signal is too weak
to be used to find strings on the sky, it might be a useful 
signature to confirm the presence of a string if one is suspected 
in some field of view.

Before closing this section it is important to point out that
our estimate treats all kinks equally whereas Copeland and Kibble 
\cite{Copeland:2009dk} emphasize that most kinks are nearly straight
and it is relevant to consider the sharpness of the kink. Hence
a more rigorous estimate should fold in the kink sharpness distribution 
with the photon emission rate, and also include effects of kink 
smoothing. (The analysis in \cite{Copeland:2009dk} accounts for
string straightening due to Hubble expansion but not kink smoothing
due to radiation backreaction.) We leave more realistic calculations 
of observational signatures for future work.

\section{Conclusions}
\label{conclusions}

We have studied the problem of an oscillating solenoid in vacuum
and determined the rate of scalar particle creation due to the 
Aharonov-Bohm interaction. This is novel because the electromagnetic
fields of a moving solenoid vanish everywhere except at the location
of the solenoid. In particular, a moving solenoid does not emit 
electromagnetic radiation but does emit charged particle pairs
as AB radiation. We have
also determined the pattern of AB radiation and find that it is
pre-dominantly in the direction perpendicular to the plane of 
oscillation.

Cosmic string loops that have AB interactions with quantum fields
will also emit AB radiation. In the thin string limit, loops 
with cusps and kinks emit a linearly divergent power. Imposing 
a cutoff due to the thickness of the string indicates that the 
AB radiation is significantly stronger than the naive dimensional
estimate.

We point out that all quantum fields interact with cosmic strings
via the gravitational AB effect, and thus there is corresponding 
gravitational AB radiation of all particles, including photons. 
Cosmic strings, even if occurring in dark matter sectors of 
particle physics, will emit photons. We have calculated the 
energy emitted in photons from loops with kinks and cusps. As 
in the gauge case, the radiated energy is much larger than the 
naive dimensional estimate. We have made a rough estimate 
of the energy emitted in photons and it remains to be seen if the
signal is strong enough to make it an interesting observational 
tool.

\begin{acknowledgments}
We would especially like to thank Juan Maldacena for many crucial 
discussions, and are grateful to Mark Alford, Thibault Damour, 
Larry Ford, Jaume Garriga, Zohar Komargodski, Michele Papucci, 
David Shih, Daniele Steer, Phil Taylor and Edward Witten for helpful 
comments. This work was supported by the U.S. Department of Energy 
at Case Western Reserve University. TV was also supported by grant 
number DE-FG02-90ER40542 at the Institute for Advanced Study. 
\end{acknowledgments}

\appendix 

\section{Moving Frame Perturbation Theory}
\label{appendixframe}

Consider a system of
coupled oscillators governed by the time-dependent Lagrangian 
\begin{equation}
L = \frac{1}{2} \sum_i \left( \frac{d x_i}{d t} \right)^2 - 
\frac{1}{2} \sum_{ij} K_{ij} (t) x_i x_j 
\label{eq:coupledoscillators}
\end{equation}
The time-dependence enters through the coupling matrix $K(t)$. 
It is helpful to determine the instantaneous normal modes 
$q_{\alpha} (t)$ with instantaneous frequencies $\omega_{\alpha} (t)$ 
that satisfy
\begin{equation}
\sum_{j}
K_{ij} (t) q_{\alpha j} (t) = \omega_{\alpha}^2 (t) q_{\alpha i} (t).
\label{eq:instmodes}
\end{equation}
These modes are normalized according to 
\begin{equation}
\sum_i q_{\alpha i} q_{\beta i} = \delta_{\alpha \beta}.
\label{eq:normalized}
\end{equation}

The oscillator co-ordinates may then be expanded 
\begin{equation}
x_i = \sum_{\alpha} \xi_{\alpha} (t) q_{\alpha i} (t)
\label{eq:normalexpansion}
\end{equation}
where $\xi_{\alpha} (t)$ are instantaneous ``normal mode'' co-ordinates. 
The Lagrangian eq (\ref{eq:coupledoscillators}) may be written in normal 
mode co-ordinates as
\begin{equation}
L = \frac{1}{2} \sum_{\beta} 
\left( \frac{d \xi_{\beta}}{d t} + A_{\beta \alpha} \xi_{\alpha} \right)
\left( \frac{d \xi_{\beta}}{d t} + A_{\beta \gamma} \xi_{\gamma} \right) -
\frac{1}{2} \sum_{\alpha} \omega_{\alpha}^2 \xi_{\alpha}^2.
\label{eq:normallagrangian}
\end{equation}
where a summation over the repeated indices $\alpha$ and $\gamma$ 
is left implicit. Here the ``transition element''
\begin{equation}
A_{\beta \alpha} (t) = \sum_{i} q_{\beta i} \frac{d q_{\alpha i}}{d t}
\label{eq:aandw}
\end{equation}
is determined by the evolution of the instantaneous normal modes. 
Eq.~(\ref{eq:normallagrangian}) generalizes the textbook treatment of 
normal modes for coupled oscillators to the case that the Lagrangian is 
time-dependent. In the limit that the coupling matrix $K$ is 
time-independent, $A = 0$, and eq (\ref{eq:normallagrangian})
reduces to the textbook normal mode Lagrangian.

From the normal mode Lagrangian eq (\ref{eq:normallagrangian}) we may 
pass to the Hamiltonian
\begin{equation}
H = \frac{1}{2} \sum_{\alpha} \left[ \Pi_{\alpha}^2 + 
\omega_{\alpha}^2 \xi_{\alpha}^2 \right]  
- \sum_{\alpha \beta} \Pi_{\alpha} A_{\alpha \beta} \xi_{\beta} 
\label{eq:hamiltonian}
\end{equation}
where the canonical momentum is
\begin{equation}
\Pi_{\alpha} = \frac{d \xi_{\alpha} }{d t} + A_{\alpha \beta} \xi_{\beta}.
\label{eq:canonicalmomentum}
\end{equation}
Hamilton's equations then read
\begin{eqnarray}
\frac{d \xi_{\alpha}}{d t} &=& 
      \Pi_{\alpha} - \sum_{\beta} A_{\alpha \beta} \xi_{\beta},
\nonumber \\
\frac{d \Pi_{\alpha}}{d t} &=& 
\sum_{\beta} \Pi_{\beta} A_{\beta \alpha} - \omega_{\alpha}^2 \xi_{\alpha}.
\label{eq:hamiltoneqs}
\end{eqnarray}
In the limit that the coupling matrix $K$ is time-independent, 
$A = 0$, and the normal mode co-ordinates simply undergo the expected 
harmonic oscillations. In the  limit that the coupling matrix varies 
``slowly'', the normal modes also vary slowly, allowing us to solve 
Hamilton's equations by treating $A$ as a perturbation.

To zeroth order the solution is 
\begin{eqnarray}
\xi_{\alpha}^{(0)} (t)  &=& \xi_{\alpha}(-T) \cos [ \omega_{\alpha} (t + T) ] 
        \nonumber \\
&+& \frac{\Pi_{\alpha}(-T)}{\omega_{\alpha}} \sin [ \omega_{\alpha} (t + T) ],
\nonumber \\
\Pi_{\alpha}^{(0)} (t) &=& \Pi_{\alpha}(-T) \cos[ \omega_{\alpha} (t + T) ] 
        \nonumber \\
&-& \omega_{\alpha} \xi_{\alpha}(-T) \sin [ \omega_{\alpha} (t + T) ].
\label{eq:zeroth-order}
\end{eqnarray}
The first-order correction is
\begin{eqnarray}
\xi_{\alpha}^{(1)} (t) & = & 
\int_{-T}^{t} d \tau \hspace{1mm} \xi_{\beta}^{(0)} (\tau) 
A_{\beta \alpha}(\tau) \cos [ \omega_{\alpha} (t - \tau) ] +
\nonumber \\
& + & \int_{-T}^{t} d \tau \frac{1}{\omega_{\alpha}} \Pi_{\beta}^{(0)} (\tau) 
A_{\beta \alpha} (\tau) \sin[ \omega_{\alpha} (t - \tau) ] 
\nonumber \\
\Pi^{(1)}_{\alpha} (t) & = & \int_{-T}^{t} d \tau \hspace{1mm} 
\Pi_{\beta}^{(0)} (\tau) A_{\beta \alpha} (\tau) 
\cos [ \omega_{\alpha} (t - \tau) ] 
\nonumber \\
& - & \int_{-T}^{t} d \tau \hspace{1mm} \omega_{\alpha} 
\xi_{\beta}^{(0)} (\tau) A_{\beta \alpha}(\tau) 
\sin[ \omega_{\alpha} (t - \tau) ] 
\nonumber \\
\label{eq:firstorder}
\end{eqnarray}

For simplicity in the perturbative solution Eqs. (\ref{eq:zeroth-order}) 
and (\ref{eq:firstorder}) and hereafter it is assumed the eigenfrequencies 
$\omega_{\alpha}(t)$ are independent of time although the eigenmodes 
$q_{\alpha}(t)$ may be time-dependent. It is not difficult to obtain 
the solution in the case of time-dependent eigenfrequencies but the 
formulae are more cumbersome and for the application to solenoids
the special formulae given here suffice. This is because as the 
solenoid moves the eigenmodes shift but their frequencies do not change. 

Thus far we have focussed on the solution to the classical equations 
of motion. However for a linear system such as 
Eq.~(\ref{eq:coupledoscillators}) the same solution applies in quantum 
mechanics if we interpret $\Pi_{\alpha}(t)$ and $\xi_{\alpha}(t)$ as 
operators in the Heisenberg picture that are solutions to the Heisenberg 
equations of motion. 

To further analyze the quantum mechanics we introduce the ladder operators
\begin{eqnarray}
a_{\alpha} (t) & = & \sqrt{ \frac{\omega_{\alpha}}{2}} \xi_{\alpha} (t) + 
\frac{i}{\sqrt{2 \omega_{\alpha}}} \Pi_{\alpha}(t)
\nonumber \\
a_{\alpha}^{\dagger} (t) & = & 
\sqrt{ \frac{\omega_{\alpha}}{2}} \xi_{\alpha} (t) - 
\frac{i}{\sqrt{2 \omega_{\alpha}}} \Pi_{\alpha}(t).
\label{eq:ladder}
\end{eqnarray}
Eqs.~(\ref{eq:ladder}), (\ref{eq:zeroth-order}) and (\ref{eq:firstorder})
show that the creation and annihilation operators at time $t$ are related 
to those at the earlier time $-T$ via the canonical transformation
\begin{eqnarray}
a_{\alpha} (t) & =  &\sum_{\beta} u_{\beta \alpha} a_{\beta} (-T) +
\sum_{\beta} v_{\beta \alpha} a_{\beta}^{\dagger} (-T), 
\nonumber \\
a_{\alpha}^{\dagger} (t) & = & \sum_{\beta} u_{\beta \alpha}^{\ast} 
a_{\beta}^{\dagger} (-T) +
\sum_{\beta} v_{\beta \alpha}^{\ast} a_{\beta} (-T),
\nonumber \\
\label{eq:canonical}
\end{eqnarray}
where the Bogolyubov coefficients 
\begin{eqnarray}
u_{\beta \alpha} & = & \delta_{\beta \alpha} e^{- i \omega_{\alpha} (t + T) } 
+  \frac{1}{2} 
\left( \sqrt{ \frac{\omega_{\alpha}}{\omega_{\beta}} }  + 
\sqrt{ \frac{\omega_{\beta}}{\omega_{\alpha}} } \right)
e^{- i \omega_{\alpha} t - i \omega_{\beta} T } 
\nonumber \\
& & \times 
\int_{-T}^{t} d \tau \hspace{1mm} A_{\beta \alpha} (\tau) 
e^{ - i (\omega_{\beta} - \omega_{\alpha} ) \tau } 
\nonumber \\
v_{\beta \alpha} & = &  \frac{1}{2} 
\left( \sqrt{ \frac{\omega_{\alpha}}{\omega_{\beta}} }  - 
\sqrt{ \frac{\omega_{\beta}}{\omega_{\alpha}} } \right)
e^{- i \omega_{\alpha} t + i \omega_{\beta} T } 
\nonumber \\
& & \times
\int_{-T}^{t} d \tau \hspace{1mm} A_{\beta \alpha} (\tau) 
e^{ - i (\omega_{\beta} + \omega_{\alpha} ) \tau }. 
\label{eq:coefficient}
\end{eqnarray}
Eqs.~(\ref{eq:canonical}) and (\ref{eq:coefficient}) contain complete 
information about quantum particle production via the dynamical 
Aharonov-Bohm effect. For example the number of particles in mode 
$\alpha$ at time $T$, 
$\langle n_{\alpha} (T) \rangle = 
\langle a_{\alpha}^{\dagger} (T) a_{\alpha}(T) \rangle$
is given by
\begin{eqnarray}
\sum_{\beta} \frac{1}{4} \left( 
\sqrt{ \frac{ \omega_{\alpha} }{ \omega_{\beta} } } - 
\sqrt{ \frac{ \omega_{\beta} }{ \omega_{\alpha} } } \right)^2 
\left |
\int_{-T}^{T} d \tau \hspace{1mm} A_{\beta \alpha} (\tau) 
e^{ - i (\omega_{\alpha} + \omega_{\beta} )\tau } 
\right | ^2 
\nonumber
\end{eqnarray}
If we take the transition matrix element to oscillate in time with frequency
$\Omega$, 
\begin{equation}
A_{\beta \alpha} (\tau) = a_{\beta \alpha} \cos \Omega \tau 
\label{eq:oscillatorytransition}
\end{equation}
then the rate of particle production in mode $\alpha$ 
\begin{equation} 
R_{\alpha} = \frac{\pi}{8} \sum_{\beta} 
\left( \sqrt{ \frac{ \omega_{\alpha} }{ \omega_{\beta} } } - 
\sqrt{ \frac{ \omega_{\beta} }{ \omega_{\alpha} } } \right)^2 
\mid a_{\beta \alpha} \mid^2 
\delta [ \Omega - ( \omega_{\alpha} + \omega_{\beta} ) ].
\label{eq:rate}
\end{equation}
Here we have made use of the usual golden rule prescription
$T = \pi \delta [ \Omega - (\omega_{\alpha} + \omega_{\beta}) ]$. 


Now let us adapt this general formalism to the case of a moving 
solenoid.
In place of Eq. (\ref{eq:coupledoscillators}) we have the Lagrangian 
\begin{equation}
{\cal L} = \mid \frac{\partial}{\partial t} \phi \mid^2 -
\mid ( \nabla - i e {\mathbf A} ) \phi \mid^2 - m^2 \mid \phi \mid^2 
\label{eq:scalarfield}
\end{equation}
that describes a massive scalar field coupled to the vector potential of 
the solenoid, which is given by Eq. (\ref{eq:singular}) 
in the singular gauge we use here.
In place of Eq. (\ref{eq:instmodes}) the instantaneous eigenmodes satisfy
\begin{equation}
- [ \nabla - i e {\mathbf A}({\mathbf r}, t) ]^2 \Psi_{\alpha}( {\mathbf r}, t)
+ m^2 \Psi_{\alpha} ({\mathbf r}, t) = \omega_{\alpha}^2 \Psi_{\alpha} 
({\mathbf r}, t).
\label{eq:scalarmodes}
\end{equation}
If the solenoid is located along the $z$-axis the eigenmodes are given 
by Eq. (\ref{eq:modes}). Here we consider a solenoid moving in the $x$-$z$ 
plane while remaining parallel to the $z$-axis; at time $t$ its 
$x$-displacement is $(v_0/\Omega) \sin \Omega t$. The eigenmodes at 
time $t$, $\Psi_{\alpha} ({\mathbf r}, t)$ are therefore obtained
by simply displacing the eigenfunctions Eq. (\ref{eq:modes}) 
along the $x$-axis by the appropriate amount. Thus the modes are 
labelled by $\alpha \rightarrow k, q, l$ and have eigenfrequency
\begin{equation}
\omega_{k,q,l} (t) = \sqrt{ k^2 + q^2 + m^2 }.
\label{eq:modefrequency}
\end{equation}
Note that for this problem the instantaneous mode frequencies do not 
vary in time although the mode functions do vary in time. 

Now the transition elements Eq. (\ref{eq:aandw}) are given by 
\begin{eqnarray}
A_{k,q,l; k',q',l'} (t) & = & \int d {\mathbf r} \hspace{1mm} 
\Psi_{k,q,l}^{\ast} ({\mathbf r}, t) 
         \frac{d}{d t} \Psi_{k',q',l'} ({\mathbf r}, t) 
\nonumber \\
& = & v_0 \cos \Omega t \int d {\mathbf r} \hspace{1mm}
\psi_{k,q,l}^{\ast} ({\mathbf r}) 
      \frac{\partial}{\partial x} \psi_{k',q',l'} ({\mathbf r}),
\nonumber 
\label{eq:transition}
\end{eqnarray}
leading to Eq.~(\ref{eq:transitionx}) for the reduced transition element 
$a_{k,q,l; k',q',l'}$ given in the main body of the paper. The explicit 
form of $a_{k,q,l; k',q',l'}$ worked out there [Eqs. (\ref{eq:transitionone})
and (\ref{eq:transitiontwo})] reveals that particles are produced only 
in modes with $l=0$ or $l=1$. 

Thus far we have focussed on classical analysis. Since this is a complex 
field to analyze the quantum field theory we must introduce two sets 
of creation and annihilation operators per mode: 
$c^{\dagger}_{k,q,l}$ and $c_{k,q,l}$ that create and annihilate particles 
in mode $k,q,l$ and $d^{\dagger}_{k,q,l}$ and $d_{k,q,l}$ that create 
and annihilate anti-particles. 

It follows from Eq.~(\ref{eq:rate}) that the rate of particle production 
in mode $(k, q, l)$ is given by
\begin{eqnarray}
R_{kql} & = & \frac{\pi}{8} \sum_{k',q',l'} 
\left( \sqrt{ \frac{ \omega_{kql} }{\omega_{k'q'l'} } } -
\sqrt{ \frac{ \omega_{k'q'l'} }{ \omega_{kql} } } \right)^2 
\nonumber \\ & \times &
\mid a_{kql; k'q'l'} \mid^2 \delta 
\left[ \Omega - ( \omega_{kql} + \omega_{k'q'l'} ) \right] 
\label{eq:scalarrate}
\end{eqnarray}
Recall that we had discretized our modes for convenience by placing 
the scalar field in a box of dimensions $a$ and $L$. Taking the continuum 
limit $a \rightarrow \infty$ and $L \rightarrow \infty$
in Eq.~(\ref{eq:scalarrate}), summing over the azimuthal quantum number 
$l$ yields the expression for ${\cal I}( k, q )$ in Eq.~(\ref{eq:ikq}), 
the rate for production of a single particle species in modes with 
wave-vectors $k, q$. 
In taking the continuum limit we make use of the usual rules
\begin{equation}
\frac{1}{L} \sum_q \rightarrow \int_{-\infty}^{\infty} \frac{d q}{2 \pi}, 
\hspace{2mm}
\frac{1}{a} \sum_k \rightarrow \int_0^{\infty} \frac{d k}{\pi}.
\label{eq:statmechrule}
\end{equation}

\end{document}